\documentclass[twocolumn,floatfix]{revtex4}
\usepackage{graphicx}
\usepackage{dcolumn}
\usepackage{amsmath}

\usepackage{amsmath,amsfonts,amssymb}
\usepackage{wrapfig}
\usepackage{graphicx}
\usepackage{bbm}
\usepackage{graphics}

\def \beq {\begin{equation}}
\def \eeq {\end{equation}}
\def \ba {\begin{eqnarray}}
\def \ea {\end{eqnarray}}

\begin{document}
\hsize\textwidth\columnwidth\hsize\csname@twocolumnfalse
\endcsname

\title{Fast Single-Charge Sensing with an rf Quantum Point Contact}
\author{D. J. Reilly, C. M. Marcus}
\affiliation{Department of Physics, Harvard University, Cambridge, MA 02138, USA}
\author{M. P. Hanson and A. C. Gossard}
\affiliation{Department of Materials, University of California, Santa Barbara, California 93106, USA}

\begin{abstract}
We report high-bandwidth charge sensing measurements using a GaAs quantum point contact embedded in a radio frequency impedance matching circuit (rf-QPC). With the rf-QPC biased near pinch-off where it is most sensitive to charge, we demonstrate a conductance sensitivity of $5 \times 10^{-6} e^{2}/h$ $ {\mathrm{Hz}}^{-1/2}$ with a bandwidth of 8 MHz.  Single-shot readout of a proximal few-electron double quantum dot is investigated in a mode where the rf-QPC back-action is rapidly switched. 
\end{abstract}
\maketitle

Mesoscopic charge sensors such as the single-electron transistor (SET) \cite{Fulton_Dolan_PRL59} and the quantum point contact  (QPC) \cite{Field_PRL93}  are at the heart of many readout technologies for quantum information processing. As electrometers, their intrinsic sensitivity \cite{Devoret_Nature00,Korotkov_PRB99} provides efficient measurement with detector noise close to the minimum allowed by quantum mechanics \cite{Clerk_PRB03}. When combined with high-bandwidth operation, these devices are attractive for application in metrology \cite{Delsing_Nature05,Fujisawa_Science06},  single photon detection  \cite{Komiyama_Nature00}, and as non-invasive charge probes at the nanoscale \cite{Lu_Nature03,DiCarlo_PRL03,Biercuk_PRB06}.
  
Combining sensitivity with fast operation is challenging because of the large  $RC$ time of the detector resistance ($>$ 50 k$\Omega$) and shunt capacitance of wire between the cold stage of a cryostat and room-temperature electronics (hundreds of pF). Nevertheless several demonstrations of single-electron detection have been reported with bandwidths in the tens of kHz \cite{ Vandersypen_APL04, Gustavsson_PRL06, Amasha_06}. An approach \cite{Schoelkopf_Science98} that circumvents the difficulty of wiring capacitance uses an impedance matching network on resonance to transform the high resistance of the detector towards the $Z_{0}$ = 50 $\Omega$ characteristic impedance of a transmission line. Changes in device resistance modulate the reflected or transmitted \cite{Fujisawa_APL00} power of a radio frequency (rf) carrier, tuned to the resonance frequency. Application of this reflectometry technique to aluminum based SETs has demonstrated charge sensing \cite{Lu_Nature03},  near quantum-limited sensitivity \cite{Devoret_Nature00,Aassime_APL01}, and bandwidths above 100 MHz \cite{Schoelkopf_Science98}.

In this Letter, we extend rf reflectometry to a semiconductor quantum point contact. We describe the reflectometer circuit in detail and explore its performance as a fast, cryogenic charge sensor, demonstrating single-electron sensing with 8 MHz bandwidth. In the regime of operation, intrinsic shot noise of the QPC makes up approximately 80\% of the total system noise.  Previous work \cite{Qin_APL06,Muller} has demonstrated modulation of rf power by a point contact operated as a voltage-controlled resistor. Here, we demonstrate application of the QPC as a fast charge sensor by performing charge-state readout of an integrated double quantum dot \cite{Petta_Science05} with fixed total charge.  In addition, we present single-shot measurements of the double dot in a mode where the rf-QPC carrier is rapidly switched, modulating the measurement back-action on a time-scale of 50 ns.

\begin{figure}[t!!]
\begin{center}
\includegraphics[width=8.0cm]{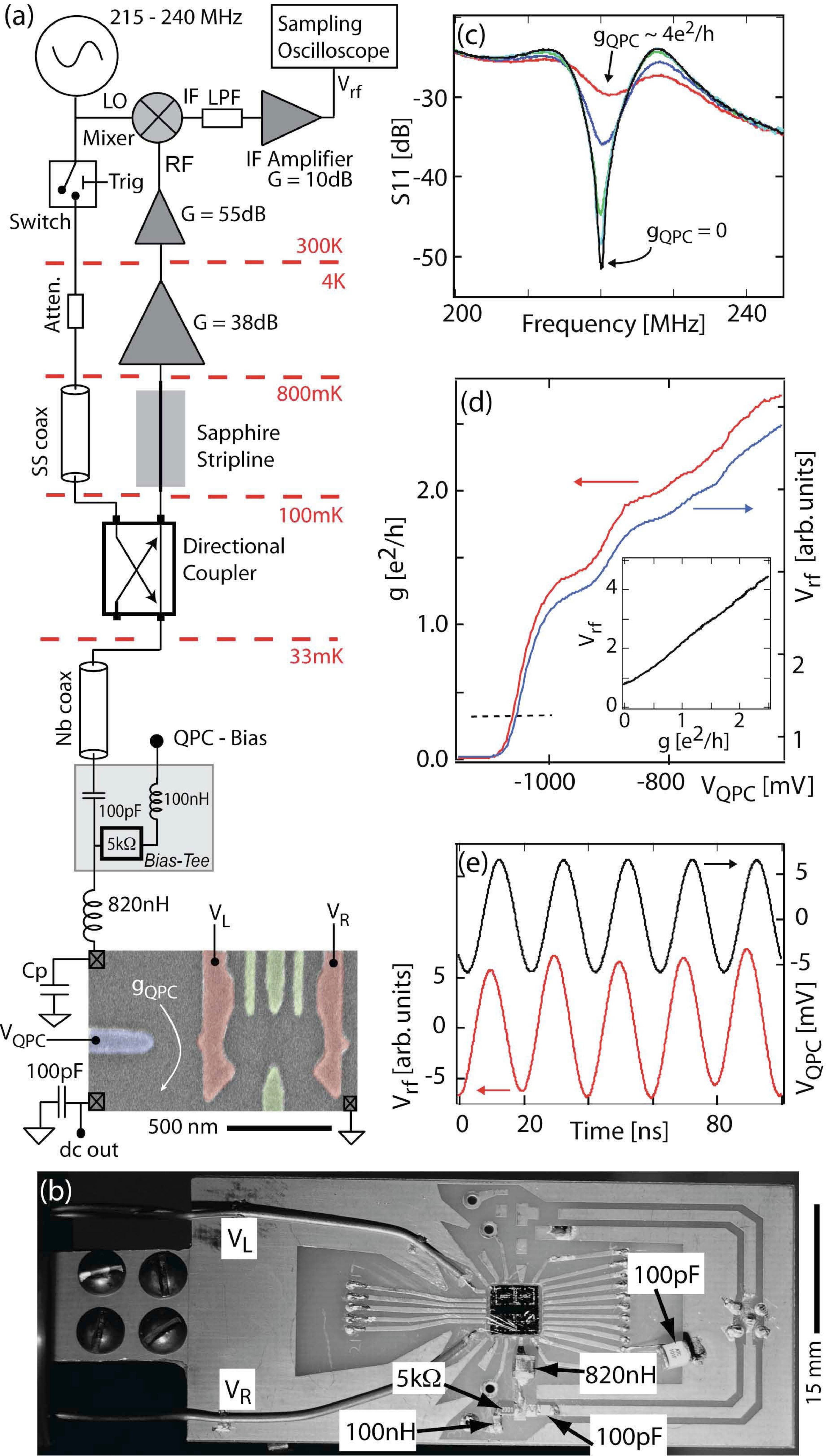}
\caption{(Color online) (a) Schematic of measurement setup including false-color SEM image of a representative device.  Gates shaded green are held at $\sim$ 1V throughout. Niobium coax and $\sim$ 50 $\Omega$ ohmic contacts minimize losses between sample and cryogenic amplifier. A stripline fabricated on a sapphire substrate is used to thermalize the inner conductor of the Nb coax. Demodulation is performed by mixing the RF signal with a local oscillator (LO) to yield an intermediate frequency (IF) that is low-pass filtered (LPF) before further amplification and digitization.  (b) Photograph of rf circuit board showing sample, matching circuit and components used to make a bias-Tee. (c) Reflection coefficient $S11$ of impedance transformer with changing $g_{QPC}$ (resonance at 220.2 MHz). (d) Demodulated response ($\mathrm{V_{ rf}}$) measured simultaneously with dc conductance as the QPC gate (shaded blue in SEM image) bias is varied ($V_{L}$ = -700mV). Dashed line indicates bias point for charge sensing. Inset shows transfer function of $\mathrm{V_{ rf}}$ verses conductance. (e) Demodulated time-domain response to 50 MHz gate signal with QPC biased to $\sim $ $e^{2}/h$ and matching inductance reduced to 560 nH (different device to all other measurements presented).
 }
\end{center}
\end{figure}

The device, shown in Fig.~1(a), consists of a
GaAs/Al$_{0.3}$Ga$_{0.7}$As heterostructure with two dimensional
electron gas (density 2 $\times$ 10 $^{15}$m$^{-2}$, mobility
20 m$^{2}$/Vs) 100 nm below the surface. Ti/Au top gates
define a few-electron double dot with proximal QPC. 

An impedance matching network is formed by a chip inductor \cite{inductor} (820 nH) and the parasitic capacitance of the bond pads and wires ($C_{p}\sim$ 0.63 pF) [see Fig.~1]. Changes in $g_{QPC}$ modulate the rf carrier power reflected from the matching network at resonance, as shown in Fig.~1(c).   Demodulation is performed at room temperature by mixing the reflected \cite{mixer} and amplified rf signal with a local oscillator (LO) (HP8647A) at the carrier frequency to yield an intermediate frequency (IF) signal, which is low-pass filtered (3dB corner frequency is 10 MHz) to remove the sum component and then further amplified (SRS SR560) to yield a voltage, $\mathrm{V_{ rf}}$, proportional to $g_{QPC}$. 

Measurements were made in a dilution refrigerator with electron temperature $T_{e} = $ 120 mK, measured by noise thermometry \cite{Spietz_APL06}. The sample is mounted on a circuit board \cite{PCB} which has 20 dc and 4 high frequency (GHz) coax lines together with 50 $\Omega$ striplines for reflectometry. The system noise temperature, $T_{N}$, determined by signal losses and the noise contribution of both the cryogenic (Quinstar) and room-temperature (Miteq AU-1565) amplifiers, was estimated to be  $T_{N} \sim$ 3.5 K by using the QPC shot noise to calibrate the gain of the system \cite{Aassime_APL01}. An on-board bias-tee enables simultaneous reflectometry and dc transport [Fig.~1(d)]. With $g_{QPC} \geq e^{2}/h$ and matching inductance reduced to 550 nH, the rf-QPC  operates in the undercoupled regime \cite{Roschier_JAP04} with a bandwidth of approximately 100 MHz determined by the rise-time of $\mathrm{V_{ rf}}$ in response to a 50 MHz bias applied to a gate, as shown in Fig.~1(e). 

\begin{figure}[t!]
\begin{center}
\includegraphics[width=8.0cm]{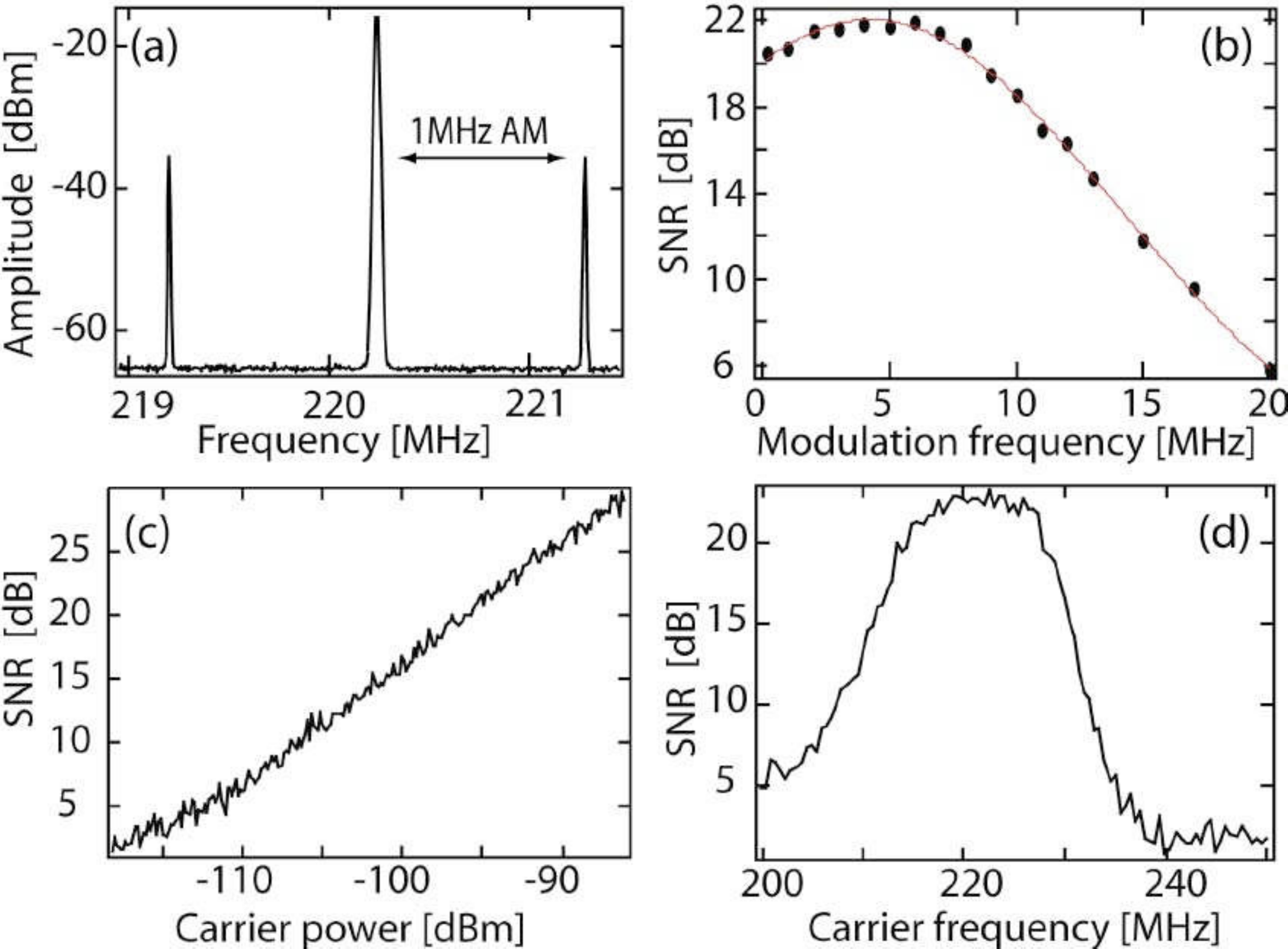}
\caption{(Color online) (a) AM response of rf-QPC to 1 MHz gate modulation, $V_{R}$ = 0.7 mV (rms). SNR of sidebands yields a sensitivity $S_g = 5\times 10^{-6} e^{2}/h$ $\mathrm{Hz^{-1/2}}$. (b) SNR of upper sideband as a function of modulation frequency. Red curve is a guide to the eye. (c) SNR of upper sideband as a function of carrier power. (d) SNR of upper sideband as a function of carrier frequency, consistent with Fig.~1(c). All SNR measurements made in a resolution bandwidth of $\Delta f = 10$ kHz.} 
\vspace{-0.5cm}
\end{center}
\end{figure}
\begin{figure}[t!]
\begin{center}
\includegraphics[width=8.0cm]{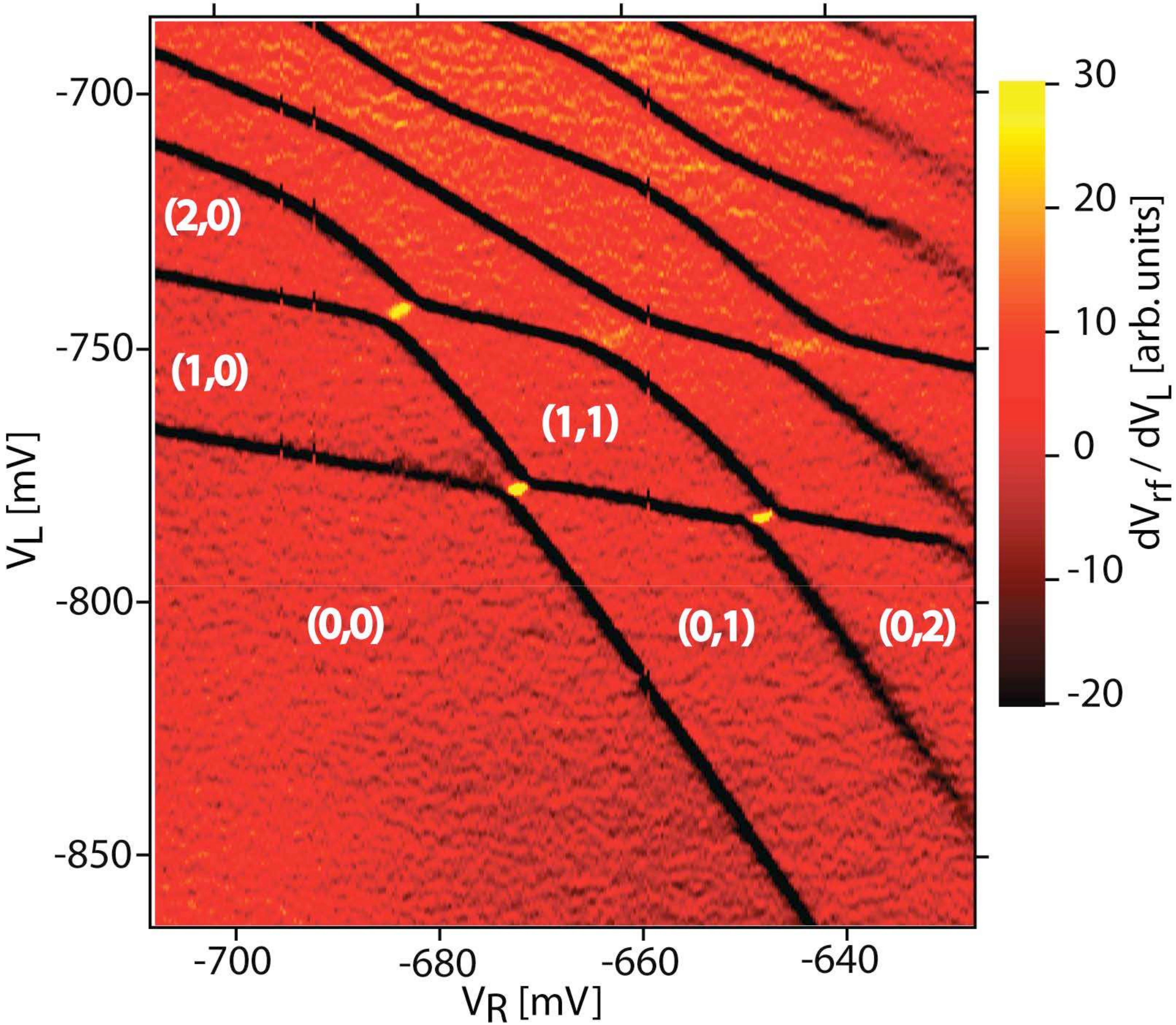}
\caption{(Color online) (a) Derivative of $\mathrm{V_{ rf}}$ (in arbitary units) as a function
of $V_{L}$ and $V_{R}$, magnetic field $B$ = 100 mT. $V_{L}$ is rastered at $\sim$ 1 mV/150 $\mu$s ($\sim$ 10$^{6}$ data points acquired in
$\sim$ 180 s). Labels indicate number of electrons in the left and right dots.} 
\vspace{-0.5cm}
\end{center}
\end{figure}
\begin{figure}[t]
\begin{center}
\includegraphics[width=8.5cm]{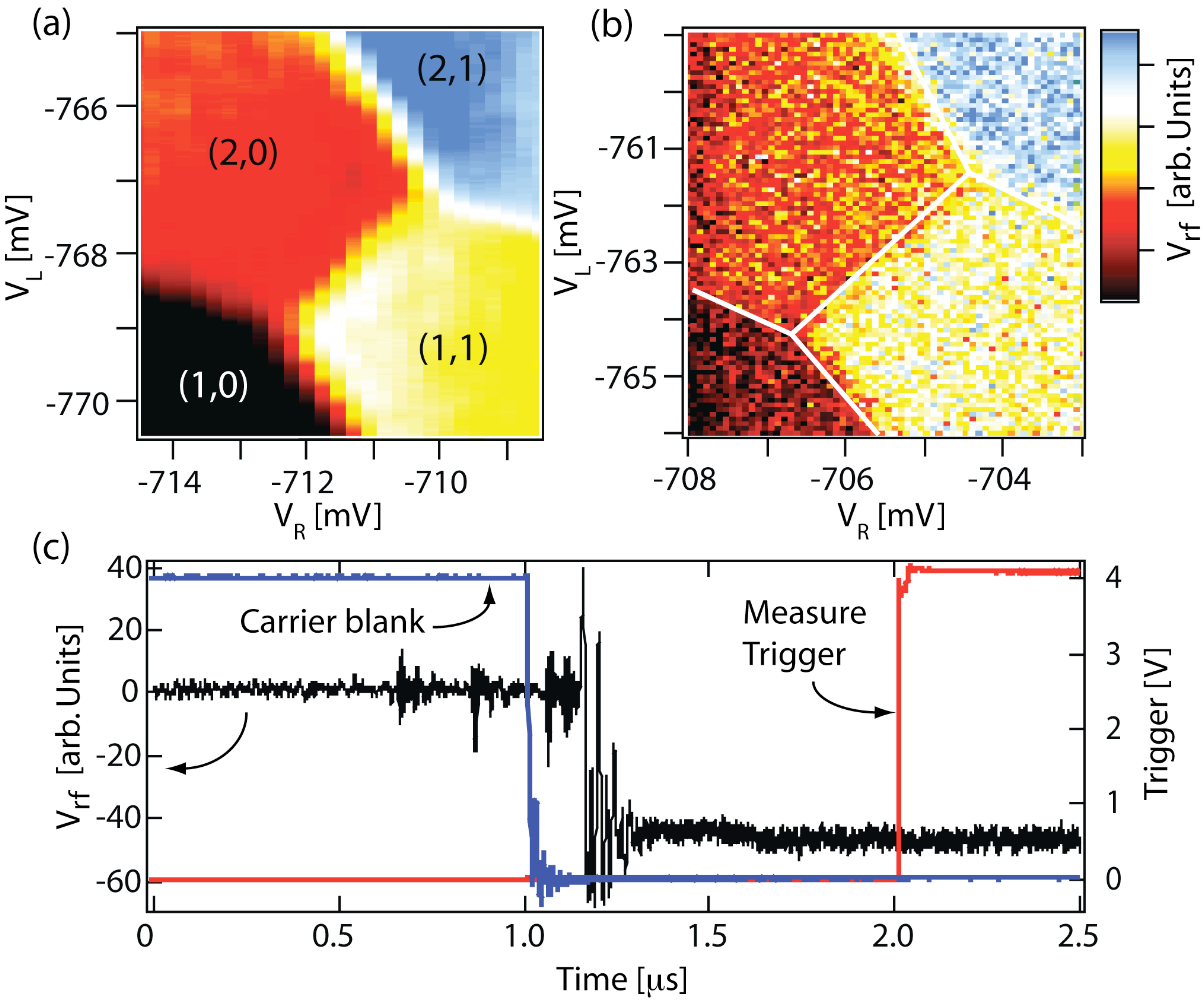}
\caption{(Color online) (a) $\mathrm{V_{ rf}}$ around the (1,1)-(2,0) transition (32 averages).  A background plane has been subtracted. (b) Single-shot readout using silent interval measurement scheme. Each pixel is an average over of an integration time $\tau_{int}$ = 60 $\mu s$. (c) Silent interval measurement scheme in which the carrier is only un-blanked during the measurement phase (blue trace). Black trace is $\mathrm{V_{ rf}}$ which is sampled for  $\tau_{int}$ = 60 $\mu s$ following the measurement trigger (red trace). An electrical delay between the room temperature un-blank trigger and $\mathrm{V_{ rf}}$ is observed.}
\vspace{-0.5cm}
\end{center}
\end{figure}

The sensitivity of the rf-QPC was determined by applying a 1 MHz 
voltage $V_{R}$ = 0.7 mV to the right gate [see Fig.~1(a)] to modulate the QPC conductance by $dg_{\rm{QPC}} \sim$ 0.018 $e^{2}/h$.  Figure 2(a) shows the resulting amplitude modulation (AM) spectrum (prior to demodulation), with sidebands symmetric about the carrier. The signal-to-noise ratio (SNR) is given by the ratio of the height of a sideband to the noise floor in a bandwidth $\Delta f$. This SNR gives a conductance sensitivity $S_g = (1/2)dg_{\rm{QPC}}(\Delta f)^{-1/2}10^{-\rm{SNR}/20}$, where the factor $1/2$ accounts for power collected from both sidebands, of $5\times 10^{-6} e^{2}/h \mathrm\, {\mathrm{Hz^{-1/2}}}$. This sensitivity allows, for instance, a $dg_{\rm{QPC}} = 0.01\,e^{2}/h$ conductance change to be measured with unity SNR in  $\tau_{int}$ = 500 ns.  Above $\sim$ 8 MHz, the $Q$-factor ($\sim$ 15) of the impedance matching circuit limits the sensitivity as shown in Fig.~2(b). The SNR increases with applied carrier power [Fig.~2(c)] up to the energy scale set by the one-dimensional sub-band spacing (typically several mV). A source-drain bias of 1 mV requires a carrier power of approximately -70dBm.  For the charge sensing measurements described below, carrier power was set to -75 dBm. For this power, $\sim$~80\% of the output noise is the intrinsic shot noise of the QPC. Figure 2(d) shows the dependence of the sideband SNR on carrier frequency, consistent with reflected power measurements [Fig.~1(c)]. 

We demonstrate the operation of the rf-QPC by detecting single-electron changes is charge configuration of a double quantum dot in the few-electron regime. For this demonstration, the QPC was biased on the steep edge of a conductance riser at $g_{QPC}\sim 0.3~e^{2}/h$, where the conductance is a sensitive function of the local electrostatic potential [see Fig.~1(d)].  Figure 3 shows ${d\mathrm{V_{ rf}}/ dV_{L}}$ as a function of gate voltages $V_{L}$ and $V_{R}$, which control the number of electrons in the left and right dots. Stable charge configurations of the double dot correspond
to the red colored regions with labels (n,m) indicating the electron occupancy on the left and right dot. Charge transitions appear in the derivative of $\mathrm{V_{ rf}}$ as black and yellow lines. 

Focusing on the (2,0)-(1,1) transition, Fig.~4(a) shows  $\mathrm{V_{ rf}}$ as a function of $V_{R}$ and $V_{L}$ with each data point averaged 32 times. In the device studied, a change in QPC conductance of $\sim 1\%$ ($\sim$ 0.003 $e^{2}/h$) is associated with an electron transition between (1,1) and (2,0). Using the measured conductance sensitivity $S_g = 5\times 10^{-6}\, e^{2}/h\, \mathrm{Hz^{-1/2}}$, we find a charge sensitivity of $\sim 10^{-3}\, e$ $\mathrm{Hz^{-1/2}}$, i.e., $\sim$ 5 $\mu$s is needed to perform charge readout with SNR of unity for this device. 

Coupling rf power to the QPC has two effects on the double dot. Firstly, an intrinsic back-action in the form of shot noise in the QPC  produces radiation that can drive transitions between double dot energy levels \cite{Ensslin_07}. Secondly, at high rf carrier power we observe some distortion in the charge stability plot [Fig.~4(a)] which is likely due to heating \cite{Vandersypen_APL04} and capacitive coupling between the oscillating QPC source-drain bias and top gates. In some applications it may be useful to eliminate the back-action of the QPC during certain phases of the duty cycle. This can be done by rapidly applying the carrier power only during a measurement using a solid-state rf-switch \cite{switch} [see Fig.~1(a)] to blank the carrier power, as shown in Fig.~4(c). The $Q$-factor of the matching network sets a time scale of $\sim$ 50 ns for the QPC to enter the measurement phase. After waiting 1 $\mu$s from the initial un-blank switch, a digitizing scope (Agilent DSO8104A) is triggered to acquire data for time $\tau_{int}$. In Fig.~4(b) we show single-shot measurements in this switched back-action mode, sampling $\mathrm{V_{ rf}}$ for $\tau_{int}$ = 60 $\mu$s during the measurement phase.  The value of each pixel is the average of $\mathrm{V_{ rf}}$ over 60 $\mu$s, which yields an SNR of $\sim$ 4, in reasonable agreement with the charge sensitivity measured at 1 MHz in the frequency domain described above.

High-fidelity readout in the present device is limited by the small coupling between the QPC and the double dot, a device parameter that can be increased considerably by improved sample design, as demonstrated, for example, in Ref.~\cite{Amasha_06}. The rf-QPC may also be useful in  detecting small changes in mesoscopic capacitance \cite{Fujisawa_APL02} that alter its resonance frequency and in the  simultaneous measurement of many rf-QPCs using multiplexing techniques \cite{stevenson_apl02, Buehler_Jap04}.

We thank  M. J. Biercuk, L. DiCarlo, E. A. Laird, D. T. McClure, and Y. Zhang for technical contributions. We especially thank J. R. Petta for experimental contributions including fabrication of the sample used. This work was supported by DARPA, DTO, NSF-NIRT (EIA-0210736), and Harvard Center for Nanoscale Systems. Research at UCSB supported in part by QuEST, an NSF Center.
\small


\begin{thebibliography}{25}
\expandafter\ifx\csname natexlab\endcsname\relax\def\natexlab#1{#1}\fi
\expandafter\ifx\csname bibnamefont\endcsname\relax
  \def\bibnamefont#1{#1}\fi
\expandafter\ifx\csname bibfnamefont\endcsname\relax
  \def\bibfnamefont#1{#1}\fi
\expandafter\ifx\csname citenamefont\endcsname\relax
  \def\citenamefont#1{#1}\fi
\expandafter\ifx\csname url\endcsname\relax
  \def\url#1{\texttt{#1}}\fi
\expandafter\ifx\csname urlprefix\endcsname\relax\def\urlprefix{URL }\fi
\providecommand{\bibinfo}[2]{#2}
\providecommand{\eprint}[2][]{\url{#2}}

\bibitem[{\citenamefont{Fulton and Dolan}(1987)}]{Fulton_Dolan_PRL59}
\bibinfo{author}{\bibfnamefont{T.~A.} \bibnamefont{Fulton}} \bibnamefont{and}
  \bibinfo{author}{\bibfnamefont{G.~J.} \bibnamefont{Dolan}},
  \bibinfo{journal}{Phys. Rev. Lett.} \textbf{\bibinfo{volume}{59}},
  \bibinfo{pages}{109} (\bibinfo{year}{1987}).

\bibitem[{\citenamefont{Field et~al.}(1993)}]{Field_PRL93}
\bibinfo{author}{\bibfnamefont{M.}~\bibnamefont{Field}} \bibnamefont{et~al.},
  \bibinfo{journal}{Phys. Rev. Lett.} \textbf{\bibinfo{volume}{70}},
  \bibinfo{pages}{1311} (\bibinfo{year}{1993}).

\bibitem[{\citenamefont{Korotkov}(1999)}]{Korotkov_PRB99}
\bibinfo{author}{\bibfnamefont{A.~N.} \bibnamefont{Korotkov}},
  \bibinfo{journal}{Phys. Rev. B.} \textbf{\bibinfo{volume}{60}},
  \bibinfo{pages}{5737} (\bibinfo{year}{1999}).

\bibitem[{\citenamefont{Devoret and Schoelkopf}(2000)}]{Devoret_Nature00}
\bibinfo{author}{\bibfnamefont{M.~H.} \bibnamefont{Devoret}} \bibnamefont{and}
  \bibinfo{author}{\bibfnamefont{R.~J.} \bibnamefont{Schoelkopf}},
  \bibinfo{journal}{Nature (London)} \textbf{\bibinfo{volume}{406}},
  \bibinfo{pages}{1039} (\bibinfo{year}{2000}).

\bibitem[{\citenamefont{Clerk et~al.}(2003)\citenamefont{Clerk, Girvin, and
  Stone}}]{Clerk_PRB03}
\bibinfo{author}{\bibfnamefont{A.~A.} \bibnamefont{Clerk}},
  \bibinfo{author}{\bibfnamefont{S.~M.} \bibnamefont{Girvin}},
  \bibnamefont{and} \bibinfo{author}{\bibfnamefont{A.~D.} \bibnamefont{Stone}},
  \bibinfo{journal}{Phys. Rev. B.} \textbf{\bibinfo{volume}{67}},
  \bibinfo{pages}{165324} (\bibinfo{year}{2003}).

\bibitem[{\citenamefont{Bylander et~al.}(2005)\citenamefont{Bylander, Duty, and
  Delsing}}]{Delsing_Nature05}
\bibinfo{author}{\bibfnamefont{J.}~\bibnamefont{Bylander}},
  \bibinfo{author}{\bibfnamefont{T.}~\bibnamefont{Duty}}, \bibnamefont{and}
  \bibinfo{author}{\bibfnamefont{P.}~\bibnamefont{Delsing}},
  \bibinfo{journal}{Nature (London)} \textbf{\bibinfo{volume}{434}},
  \bibinfo{pages}{361} (\bibinfo{year}{2005}).

\bibitem[{\citenamefont{Fujisawa et~al.}(2006)\citenamefont{Fujisawa, Hayashi,
  Tomita, and Hirayama}}]{Fujisawa_Science06}
\bibinfo{author}{\bibfnamefont{T.}~\bibnamefont{Fujisawa}},
  \bibnamefont{et~al.,}
   \bibinfo{journal}{Science} \textbf{\bibinfo{volume}{312}},
  \bibinfo{pages}{1634} (\bibinfo{year}{2006}).

\bibitem[{\citenamefont{Komiyama et~al.}(2000)\citenamefont{Komiyama, Astafiev,
  Antonov, Kutsuwa, and Hirai}}]{Komiyama_Nature00}
\bibinfo{author}{\bibfnamefont{S.}~\bibnamefont{Komiyama}},
  \bibnamefont{et~al.,}
   \bibinfo{journal}{Nature (London)} \textbf{\bibinfo{volume}{403}},
  \bibinfo{pages}{405} (\bibinfo{year}{2000}).

\bibitem[{\citenamefont{Lu et~al.}(2003)\citenamefont{Lu, Ji, Pfeiffer, West,
  and Rimburg}}]{Lu_Nature03}
\bibinfo{author}{\bibfnamefont{W.}~\bibnamefont{Lu}},
  \bibnamefont{et~al.}
  \bibinfo{journal}{Nature (London)} \textbf{\bibinfo{volume}{423}},
  \bibinfo{pages}{422} (\bibinfo{year}{2003}).

\bibitem[{\citenamefont{DiCarlo et~al.}(2004)\citenamefont{DiCarlo, Lynch,
  Johnson et~al.}}]{DiCarlo_PRL03}
\bibinfo{author}{\bibfnamefont{L.}~\bibnamefont{DiCarlo}},
  \bibnamefont{et~al.}, \bibinfo{journal}{Phys. Rev. Lett.}
  \textbf{\bibinfo{volume}{92}}, \bibinfo{pages}{226801}
  (\bibinfo{year}{2004}).

\bibitem[{\citenamefont{Biercuk et~al.}(2006)\citenamefont{Biercuk, Reilly,
  Buehler et~al.}}]{Biercuk_PRB06}
\bibinfo{author}{\bibfnamefont{M.~J.} \bibnamefont{Biercuk}},
  \bibnamefont{et~al.}, \bibinfo{journal}{Phys. Rev. B.}
  \textbf{\bibinfo{volume}{73}}, \bibinfo{pages}{201402(R)}
  (\bibinfo{year}{2006}).

\bibitem[{\citenamefont{Vandersypen et~al.}(2004)}]{Vandersypen_APL04}
\bibinfo{author}{\bibfnamefont{L.~M.~K.} \bibnamefont{Vandersypen}}
  \bibnamefont{et~al.}, \bibinfo{journal}{App. Phys. Lett.}
  \textbf{\bibinfo{volume}{85}}, \bibinfo{pages}{4394} (\bibinfo{year}{2004}).

\bibitem[{\citenamefont{Gustavsson et~al.}(2006)\citenamefont{Gustavsson,
  Leturcq, Simovic et~al.}}]{Gustavsson_PRL06}
\bibinfo{author}{\bibfnamefont{S.}~\bibnamefont{Gustavsson}},
  \bibnamefont{et~al.}, \bibinfo{journal}{Phys. Rev. Lett.}
  \textbf{\bibinfo{volume}{96}}, \bibinfo{pages}{076605}
  (\bibinfo{year}{2006}).

\bibitem[{\citenamefont{Amasha et~al.}(2006)\citenamefont{Amasha, MacLean, Radu
  et~al.}}]{Amasha_06}
\bibinfo{author}{\bibfnamefont{S.}~\bibnamefont{Amasha}},
   \bibnamefont{et~al.}, \bibinfo{journal}{(unpublished) arXiv:0607110}
  (\bibinfo{year}{2006}).

\bibitem[{\citenamefont{Schoelkopf et~al.}(1998)\citenamefont{Schoelkopf,
  Wahlgren, Kozhevnikov et~al.}}]{Schoelkopf_Science98}
\bibinfo{author}{\bibfnamefont{R.~J.} \bibnamefont{Schoelkopf}},
   \bibnamefont{et~al.}, \bibinfo{journal}{Science}
  \textbf{\bibinfo{volume}{280}}, \bibinfo{pages}{1238} (\bibinfo{year}{1998}).

\bibitem[{\citenamefont{Fujisawa and Hirayama}(2000)}]{Fujisawa_APL00}
\bibinfo{author}{\bibfnamefont{T.}~\bibnamefont{Fujisawa}} \bibnamefont{and}
  \bibinfo{author}{\bibfnamefont{Y.}~\bibnamefont{Hirayama}},
  \bibinfo{journal}{App. Phys. Lett.} \textbf{\bibinfo{volume}{77}},
  \bibinfo{pages}{543} (\bibinfo{year}{2000}).

\bibitem[{\citenamefont{Aassime et~al.}(2001)\citenamefont{Aassime, Gunnarsson,
  Bladh et~al.}}]{Aassime_APL01}
\bibinfo{author}{\bibfnamefont{A.}~\bibnamefont{Aassime}},
  \bibnamefont{et~al.}, \bibinfo{journal}{App. Phys. Lett.}
  \textbf{\bibinfo{volume}{79}}, \bibinfo{pages}{4031} (\bibinfo{year}{2001}).

\bibitem[{\citenamefont{Qin and Williams}(2006)}]{Qin_APL06}
\bibinfo{author}{\bibfnamefont{H.}~\bibnamefont{Qin}} \bibnamefont{and}
  \bibinfo{author}{\bibfnamefont{D.~A.} \bibnamefont{Williams}},
  \bibinfo{journal}{App. Phys. Lett.} \textbf{\bibinfo{volume}{88}},
  \bibinfo{pages}{203506} (\bibinfo{year}{2006}).
  
  \bibitem[{\citenamefont{Muller}(2007)\citenamefont{Muller}}]{Muller}
\bibinfo{author}{\bibfnamefont{T.}~\bibnamefont{M\"uller}},
  \bibnamefont{et~al.}, \bibinfo{journal}{28th Int. Conf. on the Physics of Semincond. Proc. AIP.},
  \textbf{\bibinfo{volume}{893}}, \bibinfo{pages}{1113} (\bibinfo{year}{2007}).

\bibitem[{\citenamefont{Petta et~al.}(2005)\citenamefont{Petta, Johnson, Taylor
  et~al.}}]{Petta_Science05}
\bibinfo{author}{\bibfnamefont{J.~R.} \bibnamefont{Petta}},
  \bibnamefont{et~al.}, \bibinfo{journal}{Science}
  \textbf{\bibinfo{volume}{309}}, \bibinfo{pages}{2180} (\bibinfo{year}{2005}).

\bibitem[{\citenamefont{designed}({\natexlab{a}})}]{inductor}
\bibinfo{author}{\bibnamefont{CoilCraft 1206CS-821XL.}}

\bibitem[{\citenamefont{designed}({\natexlab{a}})}]{mixer}
\bibinfo{author}{\bibnamefont{Mini-Circuits mixer ZP-3MH and directional coupler ZEDC-15-2B}} {\natexlab{a}}

\bibitem[{\citenamefont{Spietz et~al.}(2006)\citenamefont{Spietz, Schoelkopf,
  and Pari}}]{Spietz_APL06}
\bibinfo{author}{\bibfnamefont{L.}~\bibnamefont{Spietz}},
  \bibinfo{author}{\bibfnamefont{R.~J.} \bibnamefont{Schoelkopf}},
  \bibnamefont{and} \bibinfo{author}{\bibfnamefont{P.}~\bibnamefont{Pari}},
  \bibinfo{journal}{App. Phys. Lett.} \textbf{\bibinfo{volume}{89}},
  \bibinfo{pages}{183123} (\bibinfo{year}{2006}).

\bibitem[{\citenamefont{designed}({\natexlab{a}})}]{PCB}
\bibinfo{author}{\bibnamefont{The circuit board was designed to minimize parasitic capacitance and cross-coupling using Sonnet Suites software.}} {\natexlab{a}}

\bibitem[{\citenamefont{Roschier et~al.}(2004)\citenamefont{Roschier, Hakonen,
  Bladh, Delsing, Lehnert, Spietz, and Schoelkopf}}]{Roschier_JAP04}
\bibinfo{author}{\bibfnamefont{L.}~\bibnamefont{Roschier}},
   \bibnamefont{et~al.,}
  \bibinfo{journal}{J. App. Phys.} \textbf{\bibinfo{volume}{95}},
  \bibinfo{pages}{1274} (\bibinfo{year}{2004}).

\bibitem[{\citenamefont{Gustavsson et~al.}(2007)\citenamefont{Gustavsson,
  Studer, Leturcq, Ihn, Ensslin et~al.}}]{Ensslin_07}
\bibinfo{author}{\bibfnamefont{S.}~\bibnamefont{Gustavsson}},
  \bibnamefont{et~al.}, \bibinfo{journal}{(unpublished) arXiv:0705.3166}
  (\bibinfo{year}{2007}).

\bibitem[{\citenamefont{designed}({\natexlab{c}})}]{switch}
\bibinfo{author}{\bibnamefont{Mini-Circuits ZASWA-2-50DR.}} {\natexlab{c}}


\bibitem[{\citenamefont{Cheong et~al.}(2002)\citenamefont{Cheong, Fujisawa,
  Hayashi, Hirayama, and Jeong}}]{Fujisawa_APL02}
\bibinfo{author}{\bibfnamefont{H.~D.} \bibnamefont{Cheong}},
 \bibnamefont{et~al.,}
  \bibinfo{journal}{App. Phys. Lett.} \textbf{\bibinfo{volume}{81}},
  \bibinfo{pages}{3257} (\bibinfo{year}{2002}).

\bibitem[{\citenamefont{Stevenson et~al.}(2002)\citenamefont{Stevenson,
  Pellerano, Stahle et~al.}}]{stevenson_apl02}
\bibinfo{author}{\bibfnamefont{T.~R.} \bibnamefont{Stevenson}},
  \bibnamefont{et~al.}, \bibinfo{journal}{App. Phys. Lett.}
  \textbf{\bibinfo{volume}{80}}, \bibinfo{pages}{3012} (\bibinfo{year}{2002}).

\bibitem[{\citenamefont{Buehler et~al.}(2004)\citenamefont{Buehler, Reilly,
  Starrett et~al.}}]{Buehler_Jap04}
\bibinfo{author}{\bibfnamefont{T.~M.} \bibnamefont{Buehler}},
  \bibnamefont{et~al.}, \bibinfo{journal}{J. Appl. Phys.}
  \textbf{\bibinfo{volume}{96}}, \bibinfo{pages}{4508} (\bibinfo{year}{2004}).

\end{thebibliography}

\end{document}